\documentclass{article}
\usepackage{textcomp}

\usepackage{arxiv}

\usepackage[utf8]{inputenc} 
\usepackage[T1]{fontenc}    
\usepackage{hyperref}       
\usepackage{url}            
\usepackage{booktabs}       
\usepackage{amsfonts}       
\usepackage{nicefrac}       
\usepackage{microtype}      
\usepackage{color}

\usepackage[]{graphicx}
\usepackage{amsmath}
\usepackage{amssymb}
\usepackage{subfig}
\usepackage{caption} 
\usepackage{multicol,blindtext}
\usepackage{dblfloatfix} 
\usepackage[capitalize]{cleveref}
\usepackage{epsfig}
\usepackage{wrapfig}
\usepackage{color}

\def\##1{{\bf #1}}
\def\=#1{\underline{\underline #1}}

\def\.{\mbox{ \tiny{$^\bullet$} }}

\def\le{\left(}
\def\ri{\right)}
\def\les{\left[}
\def\ris{\right]}

\def\epso{\epsilon_{0}}
\def\lambdao{\lambda_{ 0}}
\def\muo{\mu_{ 0}}
\def\ko{k_{ 0}}
\def\etao{\eta_{0}}

\def\eps{\varepsilon}
\def\epsa{\varepsilon_a}
\def\epsb{\varepsilon_b}

\def\epsr{\varepsilon_r}
\def\epsm{\varepsilon_{met}}

\def\sp{\mathbf s}

\def\cpsi{\cos\psi}
\def\spsi{\sin\psi}

\def\ql{q^{(\ell)}}

\def\pn{^{(n)}}
\def\pom{^{(m)}}
\def\bE{{\bf E}}
\def\bH{{\bf H}}
\def\br{{\bf r}}
\def\bP{{\bf P}}
\def\bA{{\bf A}}
\def\bB{{\bf B}}

\def\ux{\hat{\#u}_x}
\def\uy{\hat{\#u}_y}
\def\uz{\hat{\#u}_z}
\def\ur{\hat{\#u}_\rho}

\title{ Surface plasmon-polariton waves obliquely guided along interface containing periodicity direction of one-dimensional photonic crystal }

\author{
Mehran Rasheed\\
Department of Physics, Lahore University of Management Sciences\\
Lahore 54792, Pakistan\\
\And
Muhammad Faryad*\\
Department of Physics, Lahore University of Management Sciences\\
 Lahore 54792, Pakistan\\
\texttt{*muhammad.faryad@lums.edu.pk}
}

\begin{document}

\maketitle

\begin{abstract}
We recently formulated the canonical boundary-value problem of propagation of surface plasmon-polariton (SPP) waves along the direction of periodicity of a one-dimensional photonic crystal. Here we present the general formulation of that canonical problem supporting the oblique propagation of SPP waves in the interface plane.  The general dispersion equation has been obtained using the rigorous coupled-wave approach for the oblique propagation and numerically solved using the Muller's method.   A periodicity in the  wavenumbers of the SPP waves was observed. Furthermore, the regions of high losses for the SPP waves, dubbed as plasmonic bandgaps, were observed in the photonic band diagram of the SPP waves. These plasmonic badngaps can be used to construct optical filters for the SPP waves.
\end{abstract}


\section{Introduction}\label{intro}
The electromagnetic surface waves at the truncated surface of a one-dimensional photonic crystal (1DPC) was systematically studied first in 1977 \cite{yeh77} and then later experimentally verified in 1978 \cite{yeh78}. These waves  were named as Bloch states or Bloch waves and often has been studied using band structures \cite{mea91}. The most popular among the category of these electromagnetic surface waves are the surface plasmon-polariton (SPP) waves \cite{polo13}. These waves exist at the interface of a dielectric material and a metal  and decay exponentially away from the either side of the interface \cite{mai07} . The SPP waves guided by the interface of a metal and 1DPC are also called Tamm plasmon-polariton waves \cite{TPP1} due to Tamm's pioneering work on the electronic surface states on the surface of a semi-infinite crystal \cite{ITamm,sho39}. 

  The 1DPCs are being utilized for optical trapping, controlling the polarization of light, and for spatial light modulation \cite{sob18}. 
The 1DPCs have also  found applications in optical biosening \cite{kon07}, utilized to couple light between alternating dielectric layers \cite{arm03}, can be used to guide modes \cite{bagh18}, and  support surface waves with backward energy flow\cite{bar08}.  Existence and excitation of the SPP waves at the interface of a metal and a 1DPC has been extensively studied \cite{arm03_1, mar06} and even multiple SPP waves at the interface perpendicular to the direction of periodicity  are well known  \cite{far10,far12}. But most of the work on 1DPCs either utilized the interface perpendicular to the direction of periodicity \cite{sob18,arm03,bagh18, bar08,arm03_1,far18} or used the band structures \cite{mor04,cag08,mou06} to obtain the surface waves. 

Recently we formulated a canonical boundary-value problem to investigate the propagation of SPP waves along the direction of periodicity of 1DPC \cite{meh18, kam19}. Here, we present a more general framework which supports the oblique propagation of the SPP waves in the interface plane that contains the direction of periodicity of the 1DPC. Because of the oblique propagation, the SPP waves are no longer $p$ polarized. The rigorous coupled wave approach (RCWA) has been used to solve the canonical problem of metal/1DPC interface \cite{Glytsis, Chateau, M.G.Moh, L.L, Faryad-1}.  

The formulation is presented in Sec. \ref{theory}. The numerical results showing the plasmonic bandgaps, power profiles, and the periodicity of the wavenumbers of the SPP waves are presented in Sec. \ref{nrd}. The concluding remarks are presented in Sec. \ref{conc}.
An $\exp(-i\omega t)$ time-dependence is implicit, with $\omega$
denoting the angular frequency. The free-space wavenumber, the
free-space wavelength, and the intrinsic impedance of free space are denoted by $\ko=\omega\sqrt{\epso\muo}$,
$\lambdao=2\pi/\ko$, and
$\etao=\sqrt{\muo/\epso}$, respectively, with $\muo$ and $\epso$ being  the permeability and permittivity of
free space. Vectors are in boldface, 
column vectors are in boldface and enclosed within square brackets, and
matrixes are underlined twice and square-bracketed. The asterisk denotes the complex conjugate, the superscript
$T$ denotes the transpose,
and the Cartesian unit vectors are
identified as $\ux$, $\uy$, and $\uz$. 

\section{Formulation of dispersion equation}\label{theory}
\begin{figure}
\begin{centering}
\includegraphics[width = 3in]{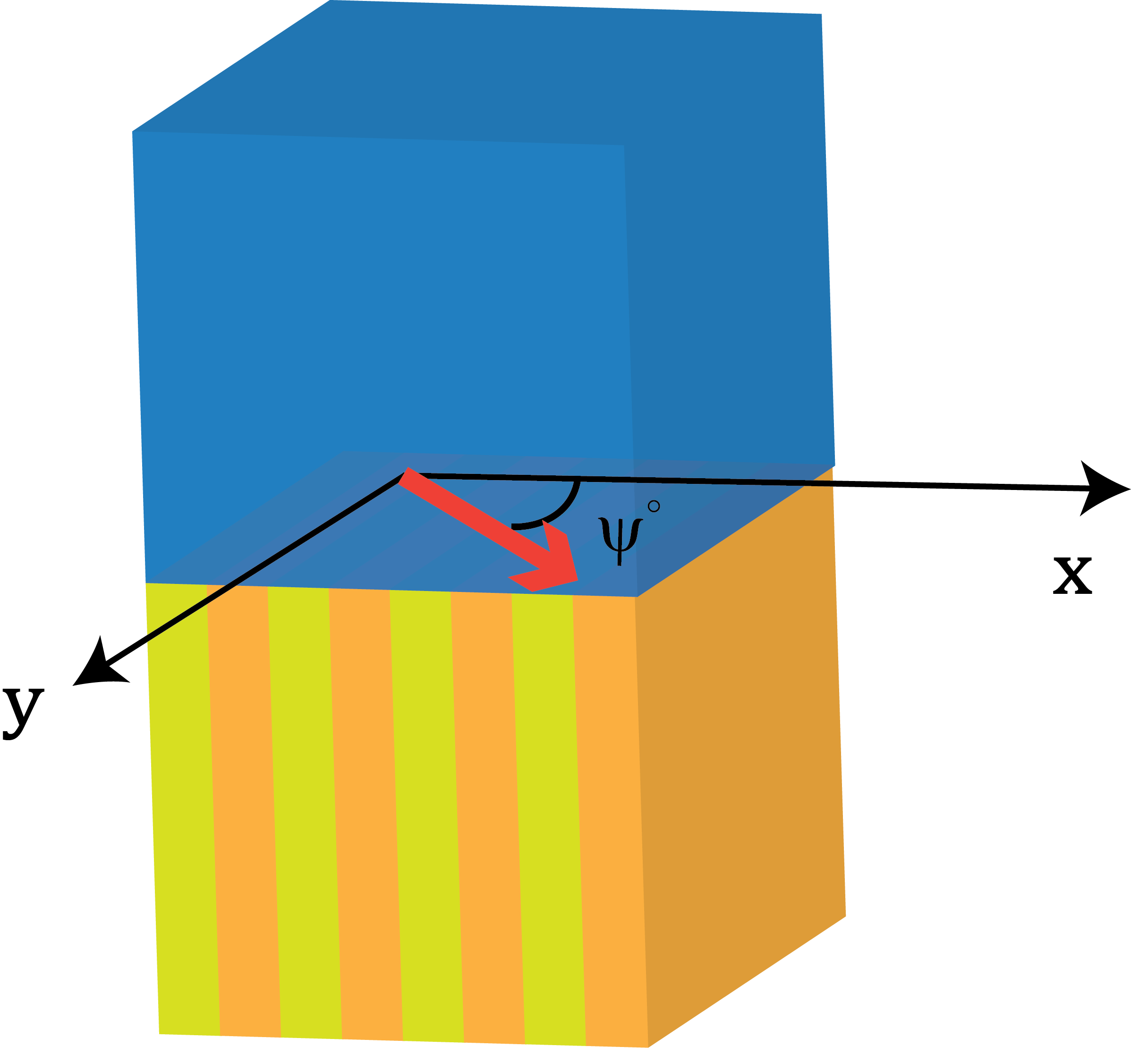}
\caption{{\bf Schematic of the canonical boundary-value problem:} The SPP waves (red, thick arrow) propagating along $\ur$ by guided by the planar interface of a semi-infinite metal ($z \leq 0$) with relative permittivity $\epsm$ and a one-dimensional photonic crystal (1DPC) occupying the half-space $z> 0$ with relative permittivity $\epsr(x) = \epsr(x \pm \Lambda)$, where $\Lambda$ is the structural period of the 1DPC.} 
\label{geom} 
\end{centering}
\end{figure}

Let us consider the boundary value problem shown schematically in Fig. \ref{geom}. The half-space $z\leq0$ is occupied by a homogeneous medium with permittivity $\epsm$ and the region $z>0$ is occupied by a one-dimensional photonic crystal of structural period $\Lambda$ along $x$ axis with the relative permittivity
\begin{equation}
\eps_r(x+\Lambda)=\eps_r(x)\,,\quad z>0\,.
\end{equation}
The relative permittivity  can be expanded as a Fourier series with respect to $x$, viz.,
\begin{equation}
\eps_r(x)=\sum_{n\in \mathbb{Z}} \eps_r^{(n)}\exp (i n2\pi x/\Lambda )\,, 
\label{perm}
\end{equation}
where
\begin{equation}
\eps_r^{(0)}=
{1\over \Lambda}\int_0^\Lambda\eps_r(x) d x\,,
\end{equation}
and
\begin{equation}
\eps_r^{(n)}=
{1\over \Lambda}\int_0^\Lambda\eps_r(x) \exp(-in2\pi x/\Lambda )d x\,,\quad n\ne0\,.
\label{permn}
\end{equation}

In the half-space $z\leq 0$, let the SPP wave propagate in the $xy$ plane making an angle $\psi$ with the $x$ axis along the unit vector
\begin{equation}
\ur = \ux \cos\psi + \uy \sin\psi\,. 
\end{equation} The field phasors may be written in terms of Floquet harmonics as follows:
\begin{eqnarray}
\bE(\br)&=&{\sum_{n\in \mathbb{Z}}}\left[\sp_n a_s^{(n)}+\#p_na_p^{(n)}\right]
\nonumber\\
&&\times\exp\left\{i\left[k_{x}^{(n)}x+k_y^{(0)}y+k_z^{(n)}z\right]\right\}\,, z\leq0\,,\label{einc}\\[5pt]
{\etao}\bH(\br)&=&  \sum_{n\in \mathbb{Z}}\sqrt{\epsm}\left[ \#p_n a_s^{(n)}-\sp_na_p^{(n)}\right]
\nonumber \\
&&\times\exp\left\{i\left[k_{x}^{(n)}x+k_y^{(0)}y+k_z^{(n)}z\right]\right\}\,, z\leq0\,,\label{hinc}
\end{eqnarray}
where  $k_x\pn=q\cos\psi+n2\pi /\Lambda$, $k_y^{(0)}=q\sin\psi$, 
and
\begin{equation}
k_z^{(n)}=
\sqrt{\epsm\ko^2-{[k_{xy}^{(n)}]}^2}\,,\label{kz}
\end{equation}
where
\begin{equation}
k_{xy}\pn=\sqrt{{[k_x\pn]}^2+{[k_y^{(0)}]}^2}\,.\label{kxy}
\end{equation}
For SPP waves propagation Im$[k_z\pn] < 0$. 
The unit vectors 
\begin{equation}
\sp_n=\frac{-k_y^{(0)}\ux +k_{x}^{(n)}\uy}{k_{xy}^{(n)}}
\end{equation}
and
\begin{equation}
\#p_n=-\frac{k_z^{(n)}}{\sqrt{\epsm}\ko}\left[\frac{k_x^{(n)}\ux +k_{y}^{(0)}\uy}{k_{xy}^{(n)}}\right]+{k_{xy}^{(n)}\over\sqrt{\epsm} \ko}\uz
\label{pn}
\end{equation}
represent the $s$- and $p$-polarization states, respectively. Whereas $\left\{a_s^{(n)},a_p^{(s)}\right\}$,
${n\in \mathbb{Z}}$, are the unknown amplitudes
of the  electric field phasor that have to be determined.

The field phasors in the 1DPC may be written in terms of Floquet harmonics as 
\begin{eqnarray}
\bE(\br)&=&\displaystyle{\sum_{n\in \mathbb{Z}}} \,\les e_x\pn(z)\ux+e_y\pn(z)\uy+e_z\pn(z)\uz\ris\nonumber\\
&&\times\exp\left\{i\left[k_x^{(n)}x+k_y^{(0)}y\right]\right\}\,, z > 0\label{fieldE}\\
\bH(\br)&=&\displaystyle{\sum_{n\in \mathbb{Z}}}\,\les h_x\pn(z)\ux+h_y\pn(z)\uy+h_z\pn(z)\uz\ris\nonumber\\
&&\times\exp\left\{i\left[k_x^{(n)}x+k_y^{(0)}y\right]\right\}\,, z>0\,.\label{field}
\end{eqnarray}
Substitution of Eqs.~(\ref{perm}), (\ref{fieldE}) and (\ref{field}) in the frequency-domain Maxwell curl postulates results in a system of six algebraic equations as follows:
\begin{eqnarray}
&&\frac{d}{dz}e_x\pn(z)-k_x\pn e_z\pn(z)=\ko{\etao} h_y\pn(z)\,,\label{max1}\\
&&\frac{d}{dz}e_y\pn(z)-k_y^{(0)}e_z\pn(z)=-\ko{\etao} h_x\pn(z)\,,\label{max2}\\
&&k_x\pn e_y\pn(z)-k_y^{(0)}e_x\pn(z)=\ko{\etao} h_z\pn(z)\,,\label{max3}\\
&&\frac{d}{dz}h_x\pn(z)-k_x\pn h_z\pn(z)=-\frac{\ko}{{\etao}}\sum_{m\in \mathbb{Z}}\eps_{r}^{(n-m)}e_y\pom(z)\,,\label{max4}\\
&&\frac{d}{dz}h_y\pn(z)-k_y^{(0)}h_z\pn(z)=\frac{\ko}{{\etao}}\sum_{m\in \mathbb{Z}}\epsilon_{r}^{(n-m)}e_x\pom(z)\,,\label{max5}\\
&&k_x\pn h_y\pn(z)-k_y^{(0)}h_x\pn(z)=-\frac{\ko}{{\etao}}\sum_{m\in \mathbb{Z}}\eps_{r}^{(n-m)}e_z\pom(z)\,.\label{max6}
\end{eqnarray}
Equations~(\ref{max1})--(\ref{max6}) hold ${\forall}n\in \mathbb{Z}$.  We restrict $|n|\leq N_t$ and then define the column $(2N_t+1)$-vectors
 \begin{eqnarray}
 [\#x_\sigma(z)]&=&[x_\sigma^{(-N_t)}(z),~x_\sigma^{(-N_t+1)}(z),~...,~x_\sigma^{(0)}(z),~...,~\nonumber\\
 &&~...,~x_\sigma^{(N_t-1)}(z),~x_\sigma^{(N_t)}(z)]^T\,,
 \end{eqnarray}
 for $\#x\in\left\{\#e,\#h\right\}$ and $\sigma\in\left\{x,y,z\right\}$. Similarly, we define $(2N_t+1)\times(2N_t+1)$-matrixes
\begin{eqnarray}
[\=K_x]={\rm{diag}}[k_x\pn]\,,\qquad
[\=\eps_r]=\left[\eps_r^{(n-m)}\right]\,,
\end{eqnarray}
where ${\rm{diag}}[k_x\pn]$ is a diagonal matrix and $[\=\eps_r]$ is a Toeplitz matrix.

Equations~(\ref{max3}) and (\ref{max6}) yield
\begin{eqnarray}
\left[\#e_z(z)\right] &=&\frac{k_y^{(0)}}{\ko}\left[\=\eps_r\right]^{-1}\cdot\left[\etao\#h_x(z)\right]\nonumber\\
&&-\frac{1}{\ko}\left[\=\eps_r\right]^{-1}\cdot\left[\=K_x\right]\cdot\left[\etao\#h_y(z)\right] \\
\left[\etao\#h_z(z)\right]&=&\frac{1}{\ko}\left[\=K_x\right]\cdot\left[\#e_y(z)\right]-\frac{k_y^{(0)}}{\ko}\left[\#e_x(z)\right],
\end{eqnarray}
the use of which in Eqs.~(\ref{max1}), (\ref{max2}), (\ref{max4}) and (\ref{max5}) eliminates $e_z\pn$ and $h_z\pn\, \forall {n \in \mathbb{Z}}$, and gives the matrix equation
\begin{equation}
\frac{d}{dz}\left[\#f(z)\right]=i[\=P]\cdot\left[\#f(z)\right]\,,\qquad z>0\,,\label{mode}
\end{equation}
where the column vector $[\#f(z)]$ with $4(2N_t+1)$ components is defined as 
\begin{equation}
\left[\#f(z)\right]=\left[\left[\#e_x(z)\right]^T,~~\left[\#e_y(z)\right]^T,~~\etao\left[\#h_x(z)\right]^T,
~~\etao\left[\#h_y(z)\right]^T\right]^T
\end{equation}
and the $4(2N_t+1)\times4(2N_t+1)$-matrix $\left[\=P\right]$ is given by
\begin{equation}
\left[\=P\right]=\left[
\begin{array}{cccc}
\left[\=0\right] & \left[\=0\right] & \left[\=P_{13}\right]&\left[\=P_{14}\right]\\
\left[\=0\right] & \left[\=0\right] & \left[\=P_{23}\right] & \left[\=P_{24}\right]\\
\left[\=P_{31}\right] & \left[\=P_{32}\right] & \left[\=0\right] & \left[\=0\right]\\
\left[\=P_{41}\right] & \left[\=P_{42}\right] & \left[\=0\right] & \left[\=0\right]\,,
\end{array}
\right]\,.\label{Pmat}
\end{equation}
where $\left[\=0\right]$ is the $(2N_t+1)\times(2N_t+1)$ null matrix and
$\left[\=I\right]$ is the  $(2N_t+1)\times(2N_t+1)$ identity matrix, the non-null submatrixes on the right side
of Eq.~(\ref{Pmat}) are as follows:
\begin{eqnarray}
\left[\=P_{13}\right]&=&\frac{k_y^{(0)}}{\ko}\left[\=K_x\right]\cdot
\left[\=\eps_r\right]^{-1}\,,\\
\left[\=P_{14}\right]&=&\ko\left[\=I\right]-\frac{1}{\ko}\left[\=K_x\right]\cdot
\left[\=\eps_r\right]^{-1}\cdot\left[\=K_x\right]\,,\\
\left[\=P_{23}\right]&=&-\ko\left[\=I\right]+\frac{[ k_y^{(0)}]^2}{\ko}\left[\=\eps_r\right]^{-1}\,,\\
\left[\=P_{24}\right]&=&-\frac{k_y^{(0)}}{\ko}\left[\=\eps_r\right]^{-1}\cdot\left[\=K_x\right]\,,\\
\left[\=P_{31}\right]&=&-\frac{k_y^{(0)}}{\ko}\left[\=K_x\right]\,\\
\left[\=P_{32}\right]&=&\frac{1}{\ko}\left[\=K_x\right]^2-\ko\left[\=\eps_r\right]\,,\\
\left[\=P_{41}\right]&=&\ko\left[\=\eps_r\right]-\frac{[k_y^{(0)}]^2}{\ko}\left[\=I\right]\,,\\
\left[\=P_{42}\right]&=&\frac{k_y^{(0)}}{\ko}\left[\=K_x\right]\,.
\end{eqnarray}

The solution of Eq. (\ref{mode}) can be written as
\begin{equation}
\left[\#f(z)\right]=\exp\left\{i[\=P]z\right\}\cdot\left[\#f(0)\right]\,,\qquad z>0\,.\label{modesol}
\end{equation}
To write the field phasors of the surface waves in the photonic crystal, we need the field expressions that represent the decay of the field away from the interface when $z \rightarrow\infty$. To do that, let us assume that $[\#t\pn]$ be the eigenvectors and $\alpha\pn$ be the corresponding eigenvalues of the matrix  $\left[\=P\right]$ and are labeled such that the first $2(2N_t+1)$ eigenvalues have Im $[\alpha\pn]>0$ so that the corresponding eigenvectors represent decaying fields as $z \rightarrow\infty$. The other half eigenvalues represent the fields that grow as $z \rightarrow\infty$. Therefore, 
\begin{equation}
\left[\#f(0+)\right]=\left[[\#t^{(1)}]~[\#t^{(2)}]~...[\#t^{2(2N_t+1)]}\right]\.\left[\bB \right]\,,\label{f0p}
 \end{equation}
where 
\begin{equation}
[\bB] = \left[b_1~b_2~...~b_{2(2N_t+1)}\right]^T\,.
 \end{equation}

 Using Eqs. (\ref{einc}) and (\ref{hinc}), we can get the components of the field phasors at $z=0-$ as
 \begin{equation}
 \left.
 \begin{array}{l}
 e_x\pn(0-)=a_s\pn\#s_n\cdot\ux+a_p^{(n)}\#p_n\cdot\ux\\
 e_y\pn(0-)=a_s\pn\#s_n\cdot\uy+a_p^{(n)}\#p_n\cdot\uy\\
 \etao\,h_x\pn(0-)=  \sqrt{\epsm} [ a_s\pn\#p_n\cdot\ux-a_p^{(n)}\#s_n\cdot\ux] \\
 \etao\,h_y\pn(0-)=\sqrt{\epsm}[a_s\pn\#p_n\cdot\uy-a_p^{(n)}\#s_n\cdot\uy]
 \end{array}\right\}\,,\quad n\in\mathbf{Z}\,
 \end{equation}
that can be expressed in matrix form as
\begin{equation}
\left[\#f(0-)\right]=\left[
\begin{array}{c}
\left[\=Y_e\right]\\[5pt]
\left[\=Y_h\right] 
 \end{array}
 \right]\cdot 
 \left[\#A\right]\label{f0m}
 \end{equation}
where
\begin{eqnarray}
\left[\#A\right]&=&\big[a_s^{(-N_t)},~a_s^{(-N_t+1)},...,~a_s^{(0)},...,~a_s^{(N_t-1)},~a_s^{(N_t)},\nonumber\\
&&~a_p^{(-N_t)}, ~a_p^{(-N_t+1)},...,~a_p^{(0)},...,~a_p^{(N_t-1)},~a_p^{(N_t)}\big]^T
\end{eqnarray}
and the non-zero entries of $(4N_t+2)\times (4N_t+2)$-matrixes $\left[\=Y_{e,h}\right]$ are as follows:
\begin{eqnarray}
\left(Y_e\right)_{nm}=
\begin{cases}
&\sp_n\cdot\ux\,,\quad n=m\in [1,~2N_t+1]\,,\\
&\#p_n\cdot\uy\,,\quad n=m\in [2N_t+2,~4N_t+2]\,,\\
&\#p_n\cdot\ux\,,\quad n=m+2N_t+1\,,\\
&\sp_n\cdot\uy\,,\quad n=m-2N_t+1\,,
\end{cases}\\
\left(Y_h\right)_{nm}=
\begin{cases}
&\sqrt{\epsm}\,\#p_n\cdot\ux\,,\quad n=m\in [1,~2N_t+1]\,,\\
&-\sqrt{\epsm}\,\#s_n\cdot\uy\,,\quad n=m\in [2N_t+2,~4N_t+2]\,,\\
&-\sqrt{\epsm}\,\#s_n\cdot\ux\,,\quad n=m+2N_t+1\,,\\
&\sqrt{\epsm}\,\#p_n\cdot\uy\,,\quad n=m-2N_t+1\,.
\end{cases}\\
\end{eqnarray}

 Implementing the standard boundary conditions, $[\#f(0+)]=[\#f(0-)]$, we get
 \begin{equation}
 [\=M]\cdot \left[
\begin{array}{c}
\left[\bB\right]\\[5pt]
\left[\bA\right] 
 \end{array}
 \right]\
= [\#0]\,,\label{M0}
 \end{equation}
 where
 \begin{equation}
 [\=M]=\left[[\#t^{(1)}]~[\#t^{(2)}]~...[\#t^{2(N_t+1)}]~
\begin{array}{c}
-\left[\=Y_e\right]\\[5pt]
-\left[\=Y_h\right] 
 \end{array}
 \right]\,.
 \end{equation}
 For non-trivial solutions, 
 \begin{equation}
{\rm det} [\=M]=0 \,,\label{dispeq}
 \end{equation}
 which is the dispersion equation of SPP waves guided by the interface with 1DPC propagating obliquely in the interface plane that the contains the direction of periodicity.

\section{Numerical Results and Discussion}\label{nrd}
For the representative numerical results in this section, the 1DPC has been taken  to consist of alternating layers of different dielectric materials $\epsa$ and $\epsb$ of equal lengths $a$ and $b=a$, respectively, such that the structural period is $a + b = \Lambda$. Hence, 
\begin{equation}
\eps_r(x)=\left\{
\begin{array}{c}
\eps_a\,,~~~ 0<x<0.5\Lambda\,,\\
\eps_b\,, ~~~0.5\Lambda<x<\Lambda\,.
\end{array}
\right.
\end{equation}
Hence the Fourier coefficients of the relative  permittivitiy are 
 \begin{equation}
\begin{split}
& \eps_r^{(0)} = \frac{(\epsa + \epsb)}{2}\,,\\ 
 &\eps_r^{(n)} = \frac{i(\epsb - \epsa)}{n\pi}\,, \quad\forall\, n = \{odd\} \,,\\
 &\eps_r^{(n)} = 0\,,\quad \forall\, n = \{even\}\,.
\end{split}
 \end{equation}
\begin{figure}
\begin{centering}
\includegraphics[width = 3in]{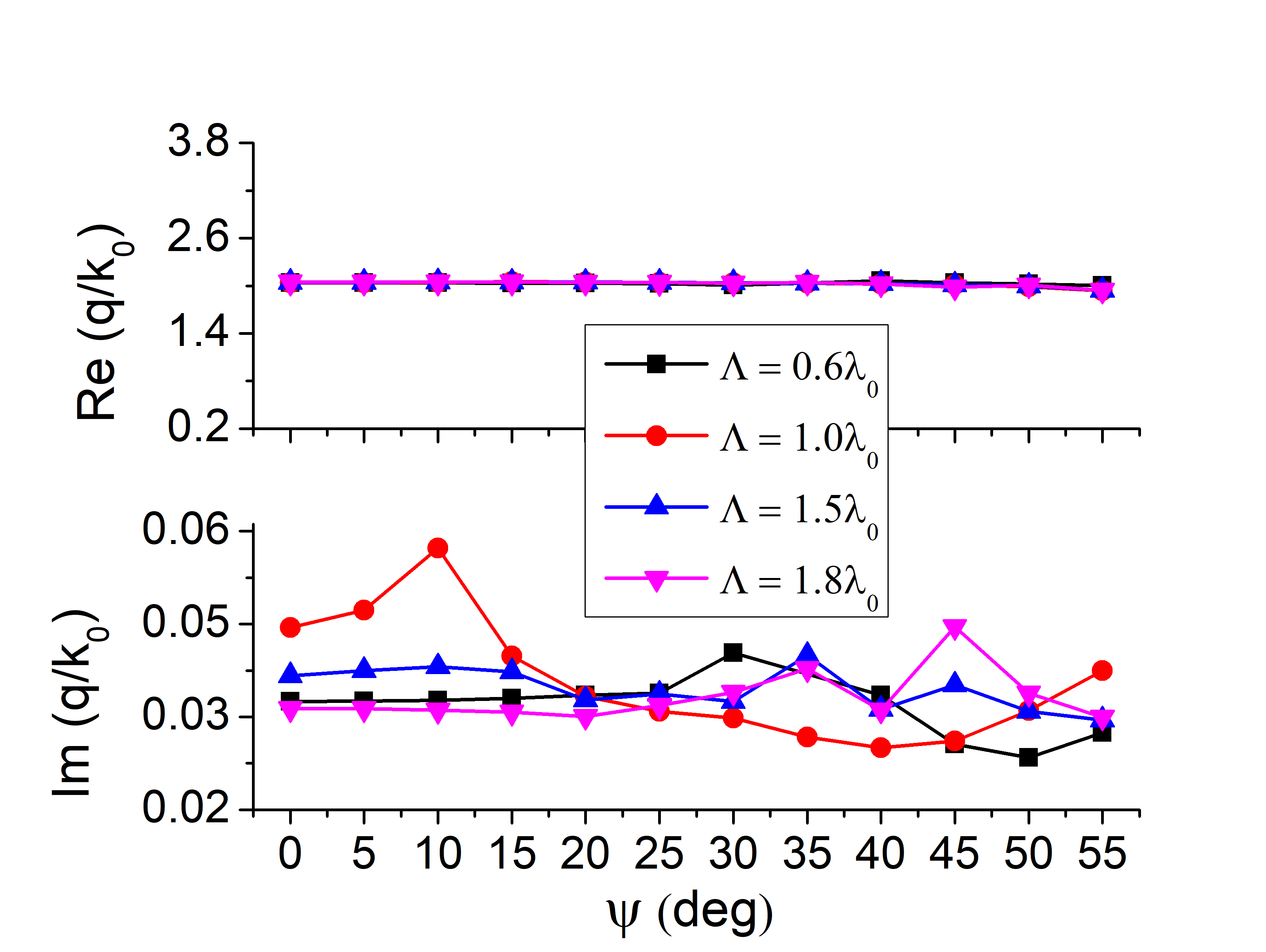}
\caption{{\bf Main branch of solutions of the dispersion equation (\ref{dispeq}):} The real and imaginary parts of the relative wavenumber $q/\ko$ as a function of the angle $\psi$ with the $x$ axis in the interface plane when the partnering metal is gold ($\epsm = -11.8+1.3i$), $\lambdao = 633 $ nm, $\epsa = (1.5)^2 + 10^{-6}i$, and $\epsb = (2)^2 + 10^{-6}i$ for different values of structural period $\Lambda$. Only the main branch of the solutions for each value of $\Lambda$ is shown.} 
\label{qmain} 
\end{centering}
\end{figure}

The relative permittvities of the layers of the 1DPC are assumed to be $\epsa = (1.5)^2 + 10^{-6}i$ and $\epsb = (2)^2 + 10^{-6}i$. The free-space wavelength of SPP waves is taken to be $\lambdao = 633$ nm and the partnering metal is chosen to be gold with $\epsm = -11.8+1.3i$ for the numerical computations. The Muller's method has been used to obtain the numerical solutions of the dispersion equation \ref{dispeq}. The relative wavenumber is first computed against $N_t$ for a fixed propagation angle so that the converged solutions can be obtained. The following relation is used as an initial guess value to obtain the main solution branch
 \begin{equation}
 q/k_0 = \sqrt{ \frac{ \epsm\eps_r^{(0)} }{ \epsm+\eps_r^{(0)} } }\,,\label{shde}
 \end{equation}
and solutions are found when the real and imaginary parts of the det$[\=M]$ were less than $10^{-6}$. Also the step-size for the Muller's method is taken to be $10^{-6}$. All the solutions of the dispersion equation (\ref{dispeq}) presented here are for $N_t=7$ and plotted against propagation angle $\psi$ with respect to the $x$ axis. 
For the chosen set of parameters,  the solutions of the dispersion equation (\ref{dispeq}) for different structural periods $\Lambda$ are presented in Fig. \ref{qmain}. Only the main branch of solutions for each value of $\Lambda$ are shown. It can be seen that real part of the relative wavenumber $q/\ko$ is almost the same for each structural period $\Lambda$, while the imaginary part of $q/\ko$ is a bit different for each $\Lambda$. Each solution presented in the \cref{qmain} represents an SPP wave obliquely propagating along the interface of 1DPC containing direction of periodicity.

\begin{figure}[h!]
\begin{centering}
\includegraphics[width = 2.8in]{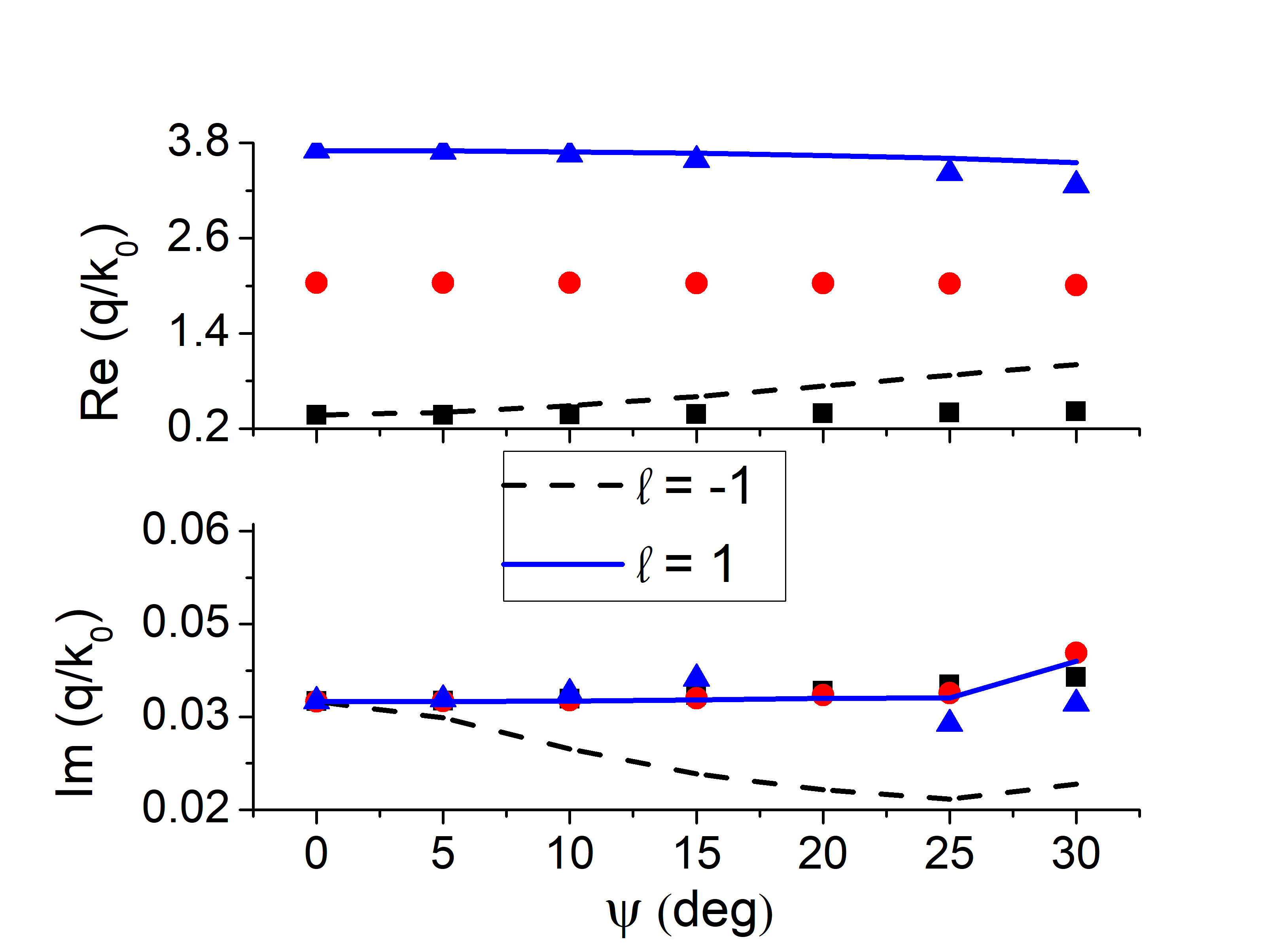}
\caption{{\bf Multiple solution for fixed $\Lambda$:} For the same parameters as Fig. \ref{qmain} except that the structural period $\Lambda = 0.6\lambdao$ is  fixed and two more branches in addition to the main branch are also presented. The symbols denote actual solution of the dispersion equation and the lines are theoretically predicted by Eq. (\ref{qell}) using the solution of the main branch. } 
\label{q06} 
\end{centering}
\end{figure}

\subsection{Periodicity of the solutions}
While solving dispersion equation (\ref{qmain}) we encountered multiple solution for a fixed $\psi$ and $\Lambda$. The multiple solutions had their imaginary part was the same but the real part was different. The three branches of these multiple solutions are presented in Fig. \ref{q06} for $\Lambda=0.6\lambdao$. The periodicity can be explained by recalling that the the wavenumber $k_{xy}\pn$ in the direction of propagation $\ur$ given by Eq. (\ref{kxy}) is valid for all the values of $|n|\leq N_t$. And, for each value of $n$, $k_{xy}\pn$ represents an evanescent Floquet harmonic  obliquely propagating along the direction of periodicity of 1DPC. Also each value of $k_{xy}\pn$ is utilized in obtaining the  dispersion equation (\ref{dispeq}). So  the solution of the dispersion equations $q$ itself possesses some sort of periodicity. We tested our hypothesis by expressing the wavenumber of the $\ell$th harmonic in Eq. (\ref{kxy}) as 
\begin{equation}
q^{(\ell)}=\sqrt{(q\cpsi+\ell2\pi/\Lambda)^2+(q\spsi)^2}\,
\end{equation}
or
 \begin{equation}
 q^{(\ell)}/{\ko} = \sqrt{\le\frac{q}{\ko}\ri^2+\le\frac{\ell\lambdao}{\Lambda}\ri^2+2\ell\frac{q\lambdao}{\ko\Lambda}\cos\psi} \,,\label{qell}
\end{equation}
using  $\ko=2\pi/\lambdao$. The plot of Eq. (\ref{qell}) for $\ell=1$ and $\ell=-1$ is provided in Fig. \ref{q06}. The very good match with the two branches indicate that these branches of solutions do not represent new SPP waves but the same as waves as the main branch.
It can be seen from  \cref{q06} that real part of the predicted and computed branches of Re$(q/\ko)$ are in better agreement  for the smaller value of the angles $\psi$ and  imaginary part of the computed and predicted values of the relative wavenumber Im$(q/\ko)$ do not agree very well for $\ell=-1$. Also the solutions of the dispersion equation (\ref{dispeq}) is really hard to find when the structural period $\Lambda \leq 1$. Let us note that the  solution obtained for $\psi=0^{\circ}$ matches exactly with the solution obtained in our previously published work \cite{meh18}.

\begin{figure}[h!]
\begin{centering}
\includegraphics[width = 2.8in]{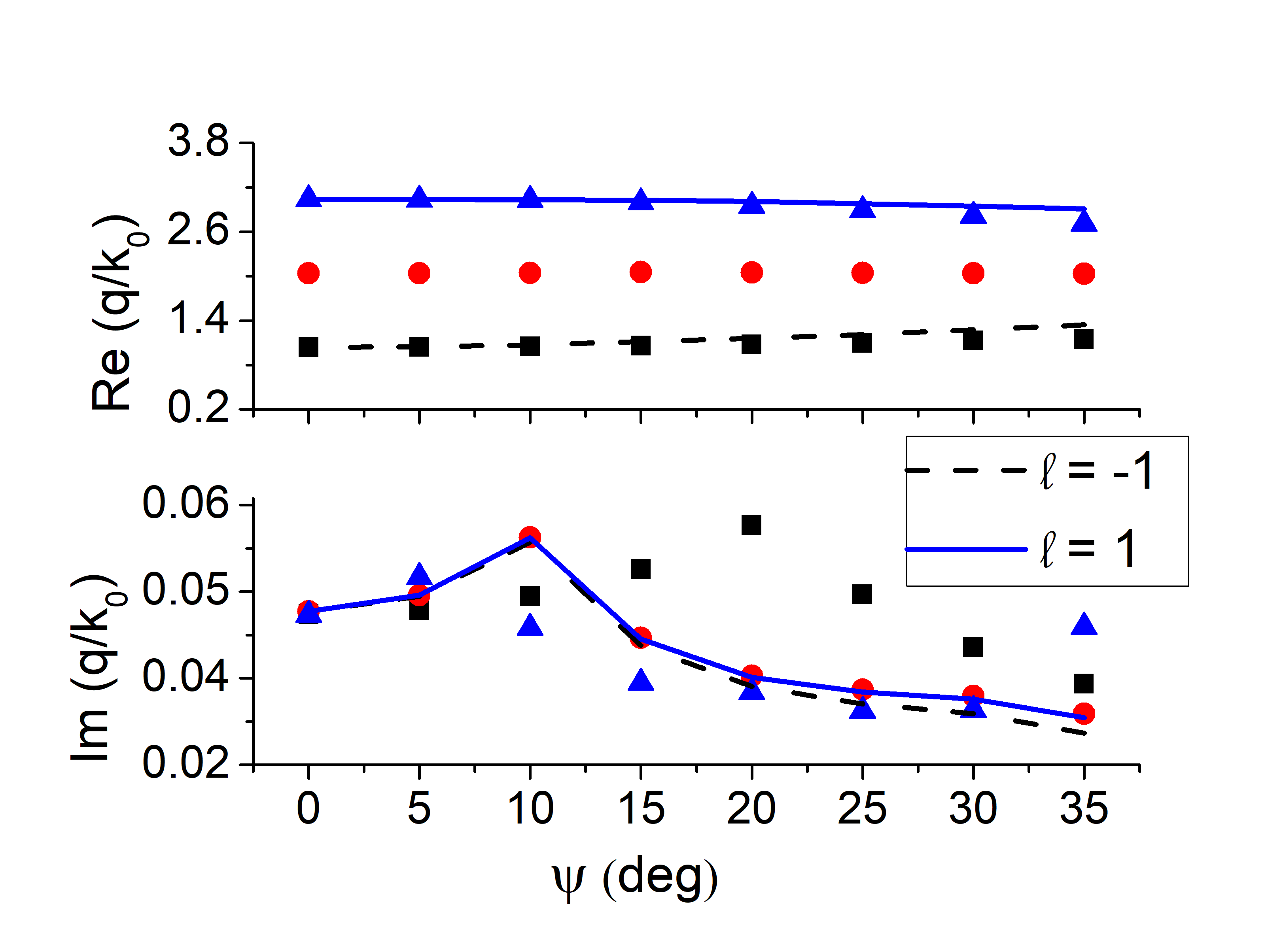}
\caption{Same  as Fig. \ref{q06} except that the structural period $\Lambda = 1.0\lambdao$ .} 
\label{q10} 
\end{centering}
\end{figure}

\begin{figure}[h!]
\begin{centering}
\includegraphics[width = 2.8in]{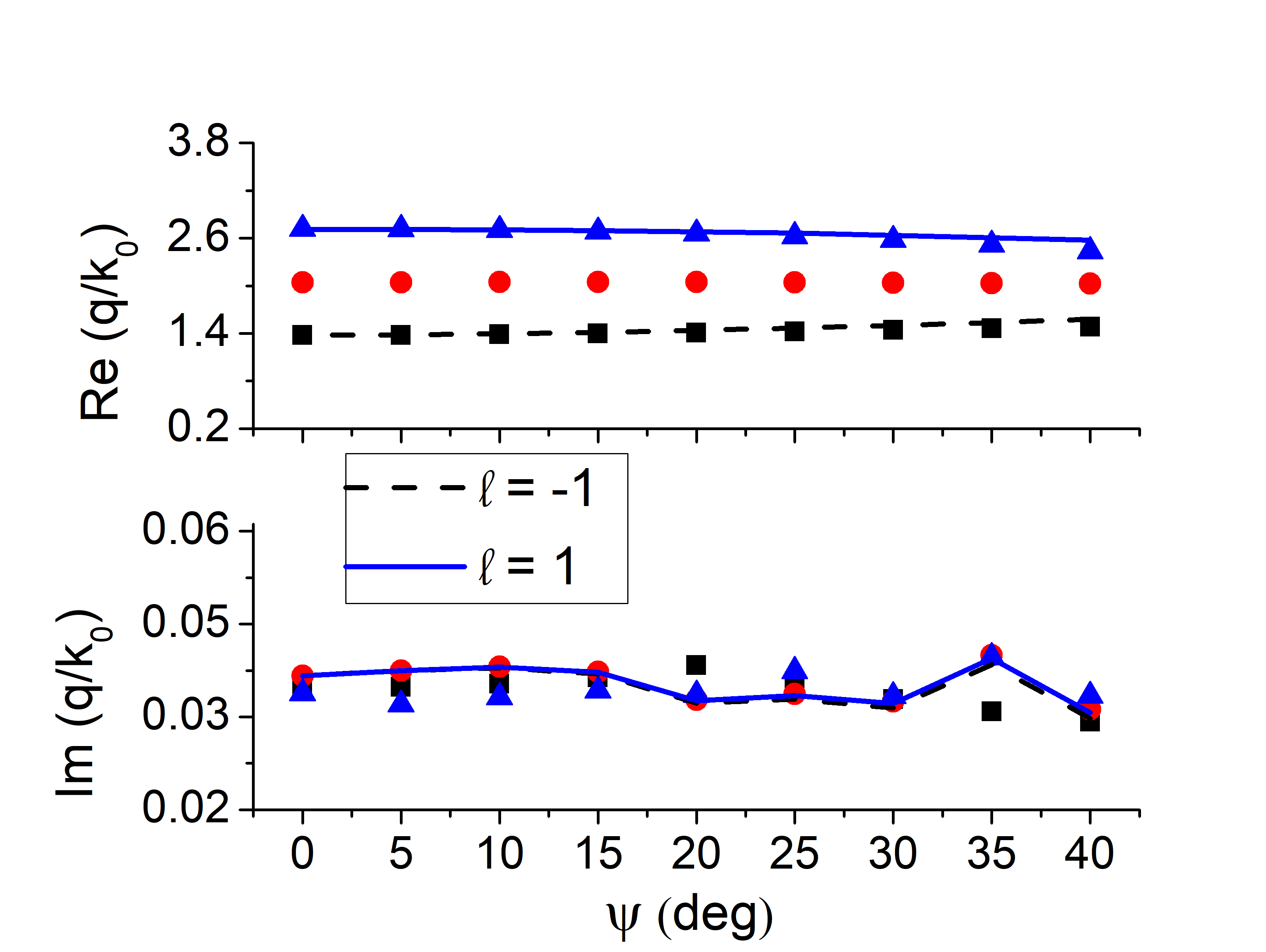}
\caption{Same  as Fig. \ref{q06} except that the structural period $\Lambda = 1.5\lambdao$ .} 
\label{q15} 
\end{centering}
\end{figure}

\begin{figure}[h!]
\begin{centering}
\includegraphics[width = 2.8in]{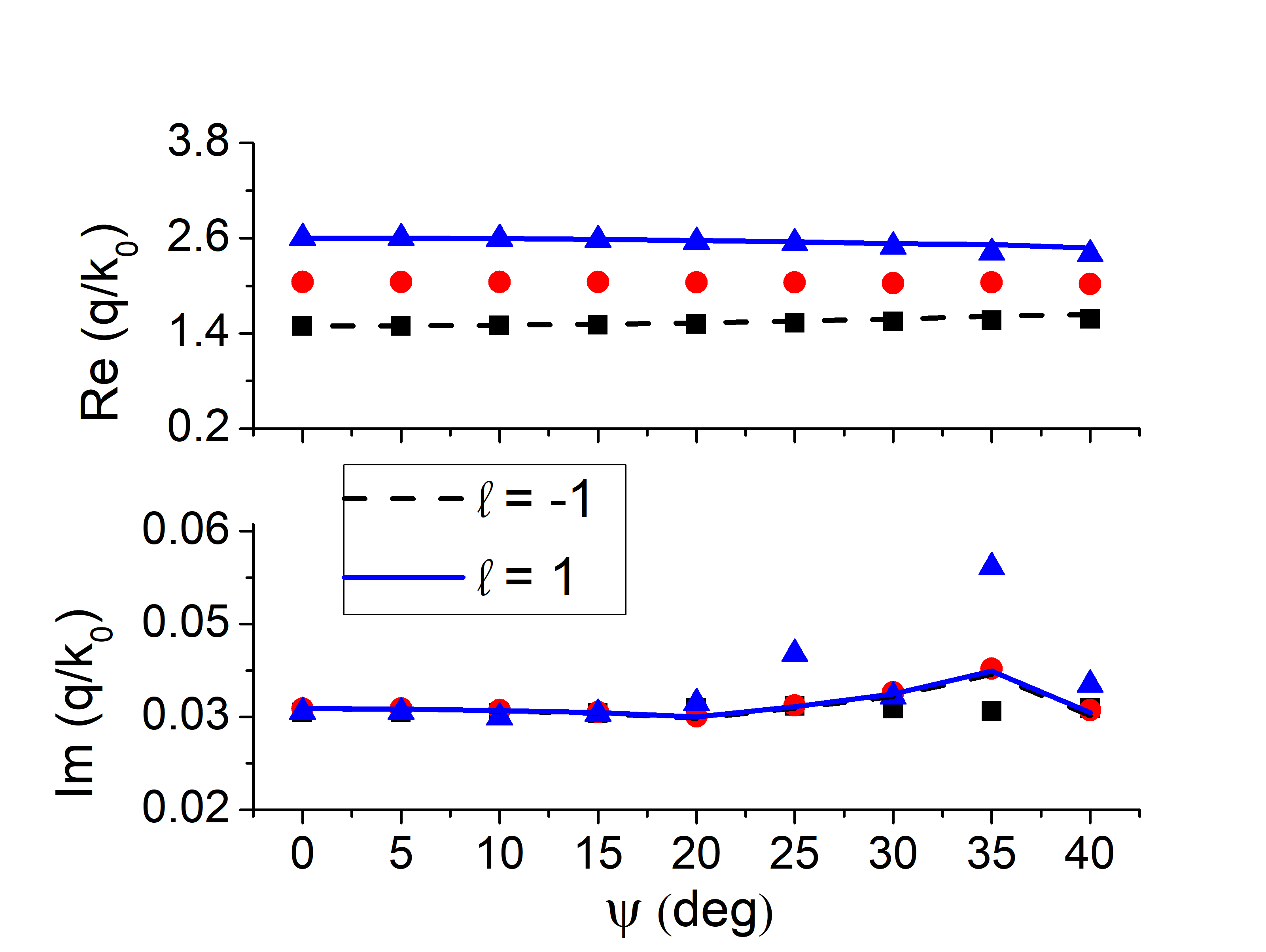}
\caption{Same as Fig. \ref{q06} except that the structural period $\Lambda = 1.8\lambdao$ .} 
\label{q18} 
\end{centering}
\end{figure}

Multiple solutions and the plots of Eq. (\ref{qell}) for $\Lambda=\lambdao$, $1.5\lambdao$, and $1.8\lambdao$ are presented in \cref{q10,q15,q18}, respectively. From these figures, we can see that the match between the computed solution of the dispersion equation and Eq. (\ref{qell}) gets better as $\Lambda$ increases. Let us not that the solutions corresponding to the higher values o $\ell$ are not shown but exist, though it gets increasing difficult to compute the solutions as $\ell$ increases in magnitude. This periodicity of the solution is just like the periodicity of the wavenumber in periodic media; however, the agreement between the solutions for the truncated 1DPC (the dispersion equation of SPP waves) and for the bulk 1DPC, i.e., Eq. (\ref{qell}), is not perfect. 

\subsection{Power profiles}
\begin{figure}[h!]
\begin{centering}
\includegraphics[width = 3.20in]{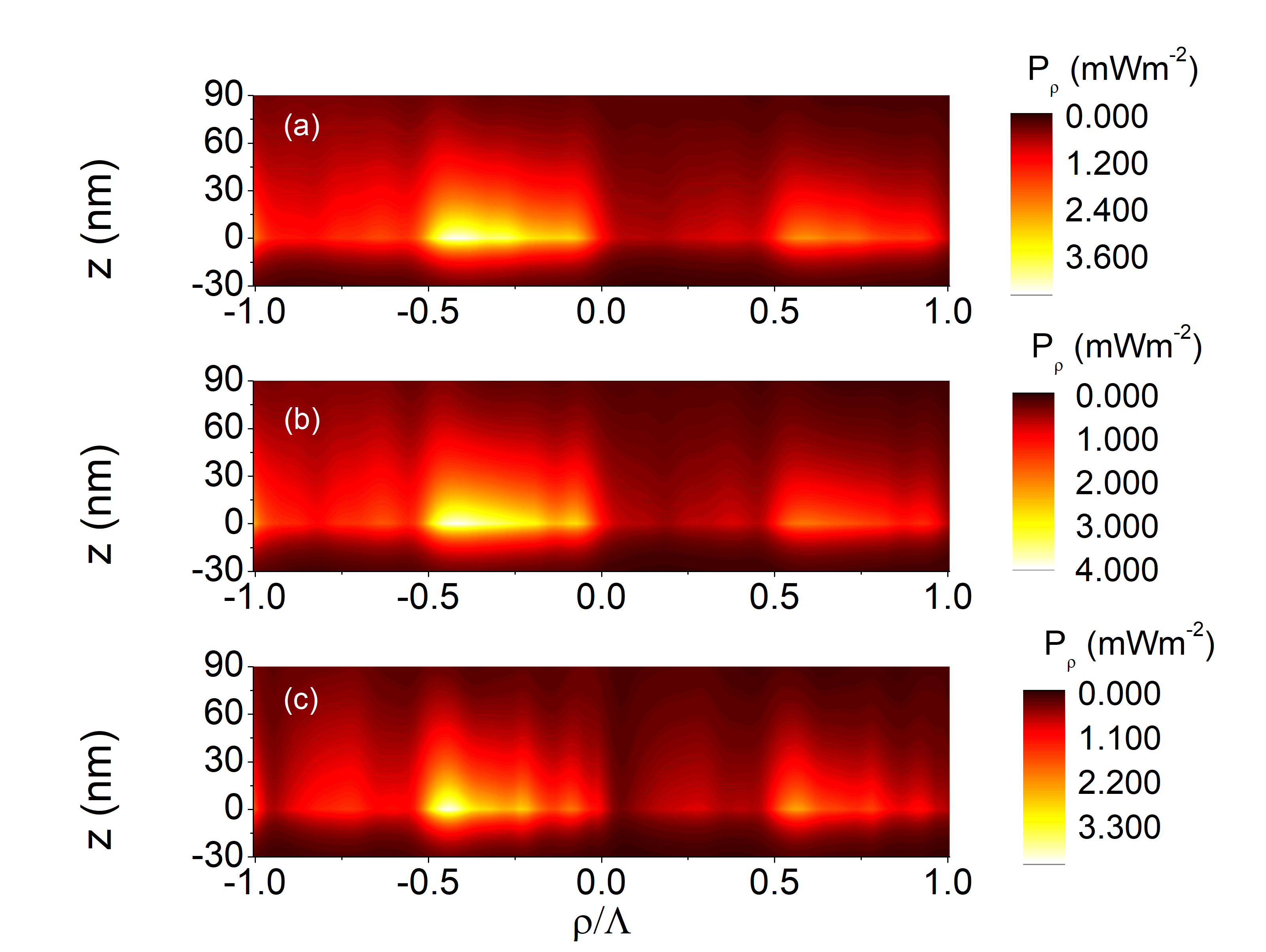}
\caption{{\bf Spatial profile of power density (\ref{dispeq}):} The valriation of the $\rho$-component of the time-averaged Poynting vector as a function of $z$ and $\rho$, when $N_t = 7$, $\psi = 10^{\circ}$, and  $\Lambda = 1.0\lambdao$  for (a) $\ell = -1$ ($q/\ko = 1.046 + 0.047i)$), (b) $\ell = 0$   ($q = 2.042+0.056i$), (c) $\ell = 1$ ($q = 3.028 +0.043i$).} 
\label{pr1} 
\end{centering}
\end{figure}

To confirm that the solutions of the dipsersion equation indeed represent SPP waves, the spatial profiles of the component of the Poynting vector along the direction of propagation $\ur$ were examined for the various solution. The representative plots are are given in \cref{pr1} for some solutions presented in \cref{q06,q10,q15,q18,qmain}. The power has been computed using the relation 
\begin{equation}
P_\rho(\ur\.\#r,z) = \ur\.\bP(\#r)\,,
\end{equation}
where
\begin{equation}
\bP(\#r) = 0.5 ~\textrm{Re} [ \bE(\#r)\times\bH^*(\#r)]\,.
\end{equation}
The power density was computed along the direction of propagation of the SPP wave. We used the $\rho$-component of the SPP wave as the SPP wave is obliquely propagating along $\ur$. Figure \ref{pr1} represents the power density of the SPP wave as a function of $\rho$ and $z$ for a fixed $\psi$ and $\Lambda$ when $\ell=\{-1,0,1\}$. It is evident the field is highly-localized to the interface and decays away from the interface. The field decays periodically along $\ur$ according to the Floquet--Lyapunov theorem \cite{YS75}. Moreover, comparison of the different graphs of the \cref{pr1} shows that the SPP waves for different values of $\ell$ are very similar, though not exactly the same. To see the effect of the direction of propagation $\psi$, the profiles for two values of $\psi$ are presented for the same $\ell$ and the same $\Lambda$. The figure shows that the field profile changes significantly when $\psi$ changes.


\begin{figure}[h!]
\begin{centering}
\includegraphics[width = 2.8in]{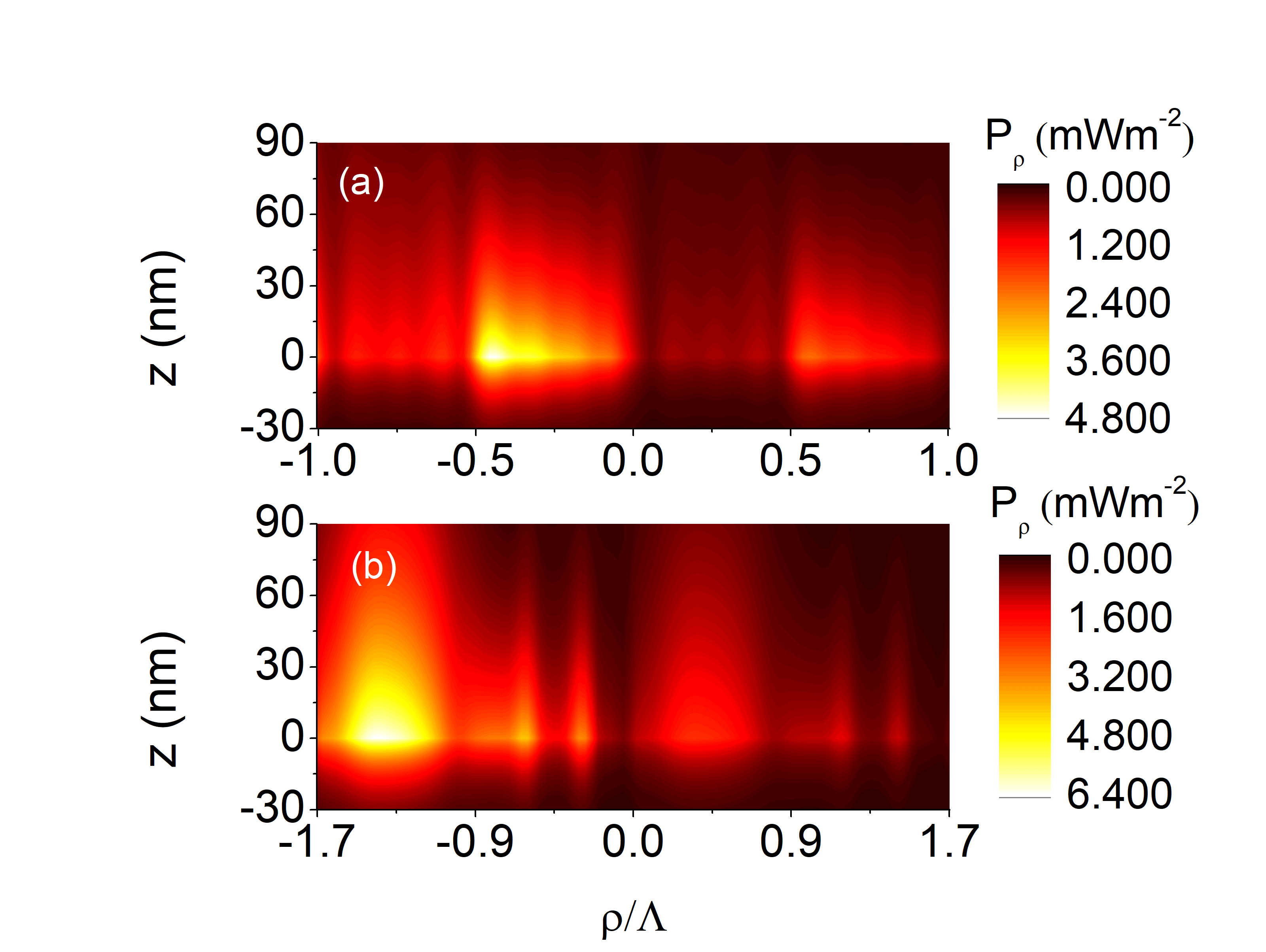}
\caption{{\bf Spatial profile of power density (\ref{dispeq}):} Same as Fig. \ref{pr1} except for $\ell = 0$, $\Lambda = 1.5\lambdao$ and (a) $\psi = 10^{\circ}$ ($q/\ko = 1.046 + 0.047i)$), (b) $\psi = 55^{\circ}$ ($q = 2.70+	0.035i$).} 
\label{pr2} 
\end{centering}
\end{figure}


\subsection{Plasmonic bandgaps for $\psi=0$ deg}
\begin{figure}[h!]
\begin{centering}
\includegraphics[width = 2.8in]{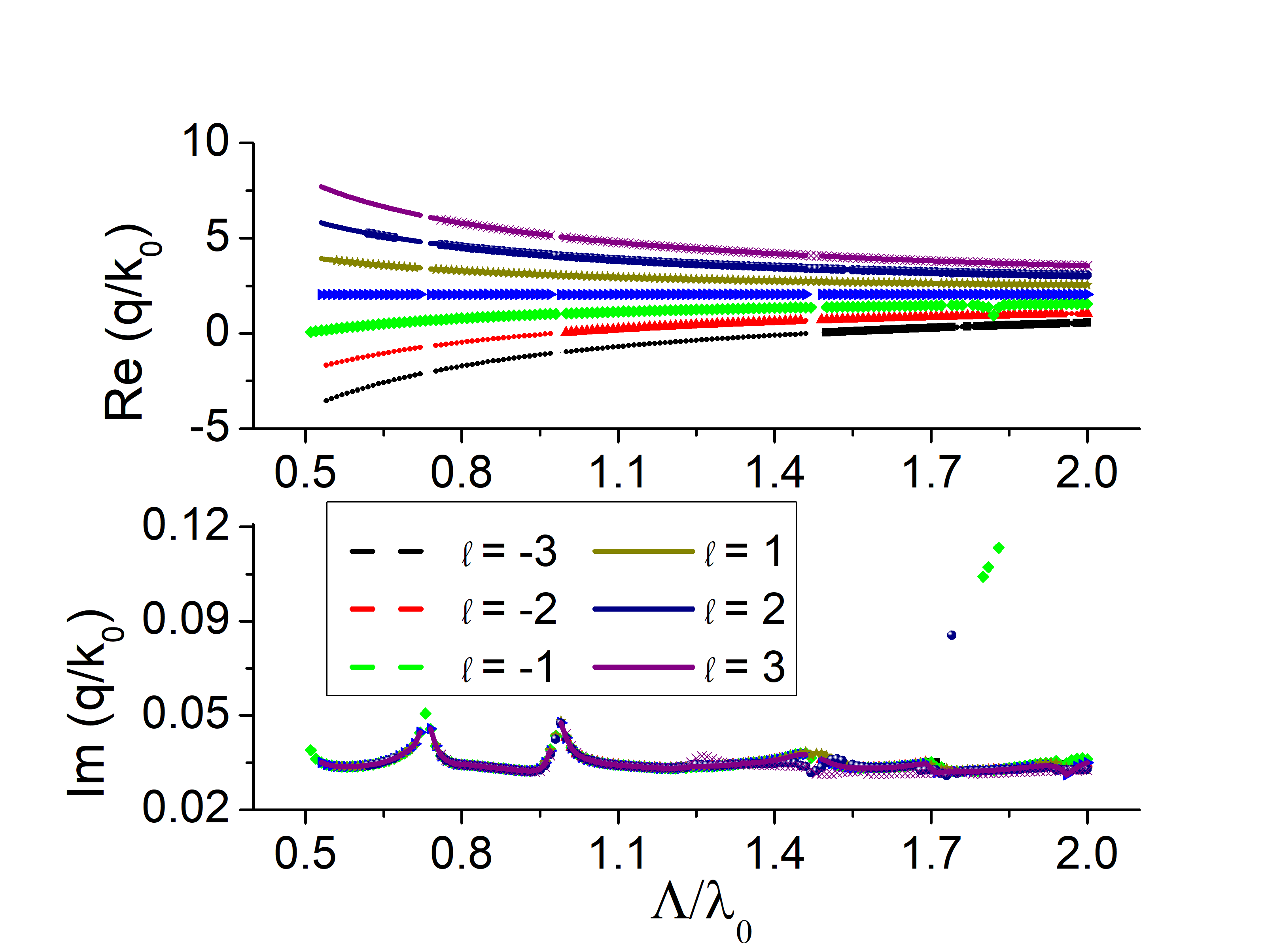}
\caption{{\bf Solution of the dispersion equation (\ref{dispeq}):} Solutions of the dispersion equation for the same parameters as Fig. \ref{qmain} except for   $\psi= 0^{\circ}$. Also plotted are the theoretically predicted upper and lower branches  using Eq. (\ref{eqq0}).} 
\label{q0} 
\end{centering}
\end{figure}

The periodicity of solutions of the dispersion equation  is further explored when the SPP wave propagates along the direction of periodicity. For $\psi=0^{\circ}$, $k_x\pn = q + n2\pi/\Lambda$ so Eq. (\ref{qell}) reduces to 
\begin{equation}
q^{(\ell)}/\ko = q^{(\ell)}/\ko + \ell\lambdao/\Lambda\,.\label{eqq0}
\end{equation} 
The solutions of the dispersion equation are presented in Fig. \ref{q0} along with the plots of Eq. (\ref{eqq0}) for different values of $\ell$. Both the solutions of the dispersion equation and the theoretical prediction match pretty well. However, there are gaps in the solutions of the dispersion equation where the imaginary part of the solution quickly grows and become much higher than the solutions for nearby values of $\Lambda$.  These plasmonic bandgaps are the regions where the SPP-wave propagation is very lossy, which is reminiscent of the photonic bandgap in the bulk 1DPC. These plasmonic bandgap can be used to do filtering of the SPP waves, just like the filtering of the waves by bulk 1DPC.

A close inspection of \cref{q0} reveals some very interesting phenomena, e.g., the computed branches are not symmetric as the Eq. (\ref{q0}) predicts. It is due to the fact that the Re$[\ql/\ko]$ becomes negative  when $\Lambda\leq1.46\lambdao$ for $\ell=-3$. Similarly, for $\ell = -2$, the Re$[\ql/\ko]$ becomes negative when $\Lambda \leq 0.97\lambdao$. We know that Re$[\ql/\ko]<0$ represents growing waves when Im$[\ql/\ko]>0$, so we computed only those wavenumbers that represents the SPP waves. 

%

\section{Concluding Remarks}\label{conc}
Using the rigorous coupled-wave approach (RCWA), a general formulation of the canonical boundary-value problem of surface plasmon-polariton (SPP) waves has been presented where the half-space is assumed to be occupied by a metal while the other half-space is assumed to be occupied by the one-dimensional photonic crystal (1DPC). A dispersion equation was obtained and numerically solved using the Muller's method.   The solutions of the dispersion equation for various values of the structural period of the 1DPC and for oblique angles of propagation were obtained. 

It was found that the spatial periodicity of the 1DPC is manifested in the periodicity of the computed wavenumbers of the SPP waves. Also, for axial propagation, plasmonic bandgaps were found where the propagation is much more lossy than at other values of the structural of period. It is hoped that the formulation presented in this work will help in ushering new applications of  in the optical filtering of the SPP waves.

\section*{Funding.}
 Higher Education Commission of Pakistan (HEC) grant NRPU $5905$.


\begin{thebibliography}{99}




\bibitem{yeh77}  P. Yeh,   A. Yariv,   and C. S. Hong, ``Electromagnetic propagation in periodic stratified media. I. General theory,''
J. Opt. Soc. Am. B  {\bf67}, 423 (1977).

\bibitem{yeh78}  P. Yeh,   A. Yariv,  and  A. Y. Cho,  ``Optical surface waves in layered media,''
 App. Phys. Lett.  {\bf 32}, 104 (1978).
 
\bibitem{mea91}  R. D. Meade,   K. D. Brommer,   A. M. Rappe,  and  J. D. Joannopoulos,  ``Electromagnetic Bloch waves at the surface of a photonic crystal,''
 Phys. Rev. B   {\bf 44}, 10961 (1991).


\bibitem{polo13}  J. A. Polo Jr., T. J. Mackay,   and A. Lakhakia, \textit{Electromagnetic Surface Waves: A Modern Perspective}  (Elsevier, 2013).


\bibitem{mai07} S. A. Maier,  
\textit{Plasmonics: Fundamentals and Applications} (Springer, 2007).

\bibitem{TPP1}
M. E. Sasin, R. P. Seisyan, M. A. Kalitteevski, S. Brand, R. A. Abram, J. M. Chamberlain, A. Yu. Egorov, A. P. Vasil{\textquotesingle}ev, V. S. Mikhrin, and A. V. Kavokin 
``Tamm plasmon polaritons: Slow and spatially compact light,''
Appl. Phys. Lett. {\bf 92}, 251112 (2008).

\bibitem{ITamm} I. Tamm, 
``$\ddot{\text{U}}$ber eine $\ddot{\text{m}}$ogliche Art der Elektronenbindung an Kristalloberfl$\ddot{\text{a}}$chen,''
{Z. Phys. A} {\bf 76}, 849--850 (1932).

\bibitem{sho39}  W. Shockley, 
``On the surface states associated with the periodic potential,''
Phys. Rev.  {\bf56}, 317 (1939).


\bibitem{sob18} I. V. Soboleva,   M. N. Romodina,   E. V. Lyubin,  and  A. A. Fedyanin,  
``Optical effects induced by Bloch surface waves in one-dimensional photonic crystals,"  App. Sci., {\bf 8}, 127 (2018).



%
%

%





\bibitem{kon07}  V. N. Konopsky  and  E. V.  Alieva,   
``Photonic crystal surface waves for optical biosensors,''
  Anal. Chem. {\bf 79}, 4729--4735 (2007).
  


  
  \bibitem{arm03} J. A. Gasper-Armenta,  F. Vila, and  T. Lop{\'e}z-R{\'\i}oz,  
``Surface waves in finite one-dimensional photonic crystals: mode coupling,''
 Opt. Commun.  {\bf 216}, 379--384 (2003).


  \bibitem{bagh18}  H. K. Baghbadorani,  D.  Aurelio,   J. Barvestani,  and M. Liscidini,  
  ``Guided modes in photonic crystal slabs supporting Bloch surface waves,''
  J. Opt. Soc. Am. B  {\bf 34}, 805--810 (2018).




\bibitem{bar08} J. Barvestani,   M. Kalafi,   A. Soltani-Vala,  and A. Namdar,  
``Backward surface electromagnetic waves in semi-infinite one-dimensional photonic crystals containing left-handed materials,''
 Phys. Rev. A {\bf 77}, 013805 (2008).
 
\bibitem{mar06} J. Martorell,  D. W. L. Sprung,  and G. V. Morozov,  
``Surface TE waves on 1D photonic crystals,''
 J.  Opt.: Pure and Applied Optics  {\bf 8}, 630--638 (2006).

     
  \bibitem{arm03_1} J. A. Gasper-Armenta and  F. Vila,  
``Photonic surface wave excitation: photonic crystal-metal interface,''
 J. Opt. Soc. Am. B  {\bf 20}, 2349--2354 (2003).
 
 \bibitem{far10}   M. Faryad  and A. Lakhtakia,  
``On surface plasmon-polariton waves guided by the interface of a metal and a rugate filter with a sinusoidal refractive-index profile,''
 J. Opt. Soc. Am. B {\bf 27}, 2218--2223 (2010).



\bibitem{far12} M. Faryad,  A. S. Hall, G. D. Barber, T. E. Mallouk, and A. Lakhtakia, 
``Excitation of multiple surface-plasmon-polariton waves guided by the periodically corrugated interface of a metal and a periodic multilayered isotropic dielectric material,''
J. Opt. Am. B {\bf 29} 704--713 (2012).

 
 \bibitem{far18} M. Faryad,
``Differentiating surface plasmon‑polariton waves and waveguide modes guided by interfaces with one‑dimensional photonic crystals,''
App. Phys. A {\bf 124}, 102 (2018).

\bibitem{mou06} R. Moussa,    Th. Koschni,    and  C. M. Soukoulis,  ``Excitation of surface waves in a photonic crystal with negative refraction: the role of surface termination,''
 Phys. Rev. B {\bf 74}, 115111 (2006).

 

\bibitem{mor04} E. Morino,   L. Mart{\'\i}n-Morino,  and  F. J. Garc{\'\i}a-Vedal,  
``Efficient coupling of light into and out of a photonic crystal waveguide via surface modes,''
 Photonic. Nanostruct.  {\bf 2}, 97--102 (2004).

\bibitem{cag08} H. Caglayan, I. Bulu, and E. Ozbay, 
``Off-axis directional beaming via photonic crystal surface modes,"
 Appl. Phys. Lett. {\bf 92}, 092114 (2008).



\bibitem{meh18} M. Rasheed and M. Faryad,
``Rigorous formulation of surface plasmon-polariton   wave propagation along the direction of periodicity of one-dimensional photonic crystal,''
J. Opt. Soc. Am. B {\bf 35}, 2957--2962 (2018). 
``Erratum,'' {\bf 36}, 1396 (2019).

\bibitem{kam19} M. Kamran and M. Faryad,
``Excitation of surface plasmon polariton waves along the direction of periodicity of a one-dimensional photonic crystal,''
Phys. Rev. A {\bf 99}, 053811 (2019).



















\bibitem{Glytsis} 
E. N. Glytsis  and T. K. Gaylord,  
``Rigorous three-dimensional coupled-wave diffraction analysis of single and cascaded anisotropic gratings,''
 J. Opt. Soc. Am. A {\bf 4}, 2061--2080 (1987).

\bibitem{Chateau}
 N. Chateau and J. P. Hugonin,   
``Algorithm for the rigorous coupled-wave analysis of grating diffraction,'' 
J. Opt. Soc. Am. A {\bf 11} ,1321--1331 (1994).

\bibitem{L.L}
 L. Li,  
``Multilayer modal method for diffraction gratings of arbitrary profile, depth, and permittivity,'' 
J. Opt. Soc. Am. A {\bf 10}, 2581--2591 (1993).

\bibitem{M.G.Moh} 
 M. G. Moharam,   E. B. Grann,  and D. A. Pommet, 
``Formulation for stable and efficient implementation of the rigorous coupled-wave analysis of binary gratings,'' 
J. Opt. Soc. Am. A {\bf 12}, 1068--1076 (1995).

\bibitem{Faryad-1}  
M. Faryad  and   A. Lakhtakia,
``Grating-coupled excitation of multiple surface plasmon-polariton waves,'' Phys. Rev. A {\bf 84}, 033852 (2011).


\bibitem{YS75}
V.A.~Yakubovich and V.M.~Starzhinskii,
{\em Linear Differential Equations with Periodic Coefficients,} Wiley (1975).






\end{thebibliography}
\end{document}